\documentclass[smallextended]{svjour3}       
\smartqed  
\usepackage{graphicx}
 \newcommand{\be}{\begin{equation}}
 \newcommand{\ee}{\end{equation}}
 \newcommand{\correct}{}
 \journalname{General Relativity and Gravitation}
\begin{document}

\title{The Schr\"{o}dinger-Newton equations beyond Newton}

\titlerunning{Schr\"{o}dinger-Newton equations beyond Newton}

\author{Giovanni Manfredi}


\institute{Giovanni Manfredi \at
              Institut de Physique et Chimie des Mat\'{e}riaux de Strasbourg, CNRS and Universit\'{e} de Strasbourg, BP 43, F-67034 Strasbourg, France \\
              Tel.: +33 3 88 10 72 14 \\
              Fax: +33 3 88 10 72 45 \\
              \email{giovanni.manfredi@ipcms.unistra.fr}
}

\date{Received: date / Accepted: date}

\maketitle

\begin{abstract}
The scope of this paper is twofold. First, we derive rigorously a low-velocity and Galilei-covariant limit of the gravitoelectromagnetic (GEM) equations. Subsequently, these reduced GEM equations are coupled to the Schr\"{o}dinger equation with gravitoelectric and gravitomagnetic potentials. The resulting extended Schr\"{o}dinger-Newton equations constitute a minimal model where the three fundamental constants of nature ($G$, $\hbar$, and $c$) appear naturally. We show that the relativistic correction coming from the gravitomagnetic potential scales as the ratio of the mass of the system to the Planck mass, and that it reinforces the standard Newtonian (gravitoelectric) attraction. The theory is further generalized to many particles through a Wigner function approach.
\end{abstract}

\maketitle

\section{Introduction}\label{sec:intro}
Recent years have witnessed a surge of interest in the so-called
Schr\"{o}dinger-Newton equations (SNEs)
\begin{eqnarray}
i\hbar \frac{\partial \Psi}{\partial t} &=& -\frac{\hbar^2}{2m} \Delta \Psi +m V({\mathbf r},t) \Psi, \label{schrodinger} \\
\Delta V &=& 4\pi G m|\Psi|^2,\label{poisson}
\end{eqnarray}
where $\Psi({\mathbf r},t)$ is the wave function and $V({\mathbf r},t)$ is
the gravitational potential in the Newtonian approximation,
$m$ is the mass of the system, $G$ is the gravitational constant, and $\hbar$ is Planck's constant. Arguably, the above set of equations constitute the simplest model where nonrelativistic quantum mechanics [Schr\"{o}dinger's equation (\ref{schrodinger})] is coupled to Newtonian gravity [Poisson's equation (\ref{poisson})]. In contrast to the standard Schr\"{o}dinger equation, the system (\ref{schrodinger})-(\ref{poisson}) is nonlinear, because the matter density $\rho({\mathbf r},t) = m|\Psi|^2$ is computed in terms of the wave function $\Psi$.

The SNEs have been proposed in various (more or less speculative) contexts.
Originally, they were put forward independently by Diosi \cite{diosi}
and Penrose \cite{penrose96,penrose98} as a fundamental modification of quantum mechanics for massy objects. The underlying idea was that a linear superposition of two quantum states would give rise to two space-time geometries, which poses serious conceptual problems from the viewpoint of general relativity \cite{penrose96}. Penrose and Diosi thus suggested that the collapse of the wave function might be related to gravitational effects, and proposed the (stationary) SNEs as a possible candidate for such gravitationally-induced collapse.

It has also been suggested that if gravity -- unlike other forces -- is not quantized \cite{carlip}, then the stress-energy tensor $T_{\mu \nu}$ in Einstein's equations should be replaced by its quantum-mechanical average $\langle T_{\mu \nu} \rangle$. Such modified Einstein's equations reduce to the SNEs in the Newtonian and low-velocity limit. Alternatively, the SNEs can be derived from an expansion in $1/c$ (where $c$ is the speed of light) of the Einstein-Klein-Gordon and Einstein-Dirac system \cite{giulini2012}.
Finally, in the astrophysical literature the SNEs have been used to study self-gravitating objects (boson stars) \cite{schunk} or to describe dark matter by means of a scalar field \cite{guzman}.

The properties of the stationary SNEs were investigated in quite some detail during the past two decades \cite{moroz,harrison,tod}. More recently, numerical simulations of the nonlinear time-dependent SNEs \cite{giulini2011,vanmeter,manfredi} revealed that gravitational effects start affecting the Schr\"{o}dinger dynamics for masses larger than a certain critical value $m_c \approx \hbar^{2/3}R^{2/3} G^{-1}$, where $R$ is the size of the object.

There has been a recent debate on the derivation and the validity of the SNEs \cite{Anastopoulos,Giulini14,Bahrami}. It is clear that they do not provide a full theory of how gravity might influence, or indeed modify, standard quantum physics. Nevertheless, as a minimal model where nonrelativistic quantum mechanics meets Newtonian gravity, the SNEs may be a useful guide to future theoretical and experimental developments.

The SNEs involve two of the fundamental constants of nature, namely the gravitational constant $G$ and Planck's action $\hbar$. Often, their results are expressed in terms of Planck's units (i.e., units derived from $G$, $\hbar$, and $c$), but this is not an appropriate system, as the speed of light appears nowhere in the SNEs. Indeed, the SNEs do not include any special-relativistic effect, and are Galilei (not Lorentz) covariant.

Is it possible to extend te SNEs to include relativistic effects, at least to some lower order? The present work attempts to answer this question. More precisely, we would like to incorporate relativistic effects without breaking the Galilei covariance of the equations. The strategy adopted here is to first replace Poisson's equation (\ref{poisson}) with the equations of gravitoelectromagnetism (GEM). These are a linearized approximation of the Einstein equations  (up to fourth order in $c^{-1}$), which is formally almost identical to the Maxwell equations of ordinary electromagnetism (EM).
The Maxwell equations possess two distinct nonrelativistic (Galilei covariant) limits, the so-called electric and magnetic limits \cite{LBLL}, of which only the former is relevant to GEM. Here, we will provide a rigorous derivation of the electric limit of the GEM equations using the methods developed in \cite{manfredi_maxw}. The resulting equations can be coupled to the Schr\"{o}dinger equation to obtain a new set of SNEs augmented by a gravitomagnetic field.
These modified SNEs, though still Galilei covariant, now incorporate all three fundamental constants ($G$, $\hbar$, and $c$) and can be conveniently expressed in Planck's units. The properties of such a set of equations will be investigated in some detail.

\section{Galilean gravitoelectromagnetism }\label{sec:gem}
When the space-time metric is almost Minkowskian and terms of order $O(c^{-4})$ or higher are neglected, it is possible to write the (linearized) equations of general relativity (GR) in a form that is almost identical to that of the Maxwell equations of ordinary electromagnetism. The equations of such ``gravitoelectromagnetism" have been discussed in many good reviews \cite{Mashhoon,Ruggiero}. Their notation is not standard and here we will adopt the same convention as in \cite{Mashhoon}, except that the sign of both the gravitoelectric and gravitomagnetic fields are reversed (in order to preserve the form of the Newtonian Poisson's equation). With this convention, the GEM equations read as:
\begin{eqnarray}
\nabla\cdot {\mathbf E} &=& - 4\pi G\rho,  \label{maxwell1}\\
\nabla\cdot {\mathbf B} &=& 0,  \\
\nabla\times {\mathbf E} &=& -\frac{1}{2c} \frac{\partial {\mathbf B}}{\partial t}, \\
\nabla\times {\mathbf B} &=&-\frac{8\pi G}{c} {\mathbf J} + \frac{2}{c}\frac{\partial {\mathbf E}}{\partial t}
\,. \label{maxwell4}
\end{eqnarray}
The EM Maxwell equations (in CGS units) are recovered by taking $G=1$ and changing the sign in front of the source terms (because gravity is attractive). In addition to these obvious differences, there is an nontrivial factor of 2 appearing in the two curl equations,
{\correct
which has the same mathematical reasons as the spin 2 of the graviton.
}

The corresponding Lorentz force per unit volume is
\be
\delta\mathbf{F}= \rho\mathbf{E}+ \frac{2}{c} \mathbf{J} \times \mathbf{B}
\label{lorentz_force}
\ee
(note another factor of 2 in front of the magnetic term). The scalar and vector potentials $(V,~ \mathbf{A})$ are defined as:
\begin{eqnarray}
 \mathbf{E} &=& -\nabla V - \frac{1}{2c} \frac{\partial \mathbf A}{\partial t}, \label{potentials_e}\\
 \mathbf{B} &=& \nabla \times \mathbf{A},
 \label{potentials}
\end{eqnarray}
and the Lorentz gauge condition is
\be \frac{\partial V}{\partial t} + \frac{c}{2}\nabla
\cdot \mathbf{A} = 0.
\label{gauge}
\ee

Now, we want to rewrite Eqs. (\ref{maxwell1})-(\ref{maxwell4}) in dimensionless form. Following the procedure developed in \cite{manfredi_maxw} for the EM Maxwell equations, we normalize space to a reference length $L$ and time to a reference time $T$, which define a typical velocity $\overline{u}=L/T$. In GEM, both fields have the dimensions of an acceleration, so we normalize $\mathbf{E}$ to $a=LT^{-2}$ and $\mathbf{B}$ to $2a$ (in order to eliminate the extra factor 2 in the GEM equations).
We further rescale the mass density to a reference value $\overline{\rho}$ and the current to $\overline{u} \overline{\rho}$.
{\correct
Concrete physical values for $L$, $T$, and $\overline{\rho}$ will be
specified in the next section.
}

In these units, the GEM equations become:
\begin{eqnarray}
\nabla\cdot {\mathbf E} &=& -\frac{\rho}{\alpha}  \\
\nabla\cdot {\mathbf B} &=& 0  \\
\nabla\times {\mathbf E} &=& - \beta \frac{\partial {\mathbf B}}{\partial t} \\
\nabla\times {\mathbf B} &=& -\frac{\beta}{\alpha} {\mathbf J} + \beta\frac{\partial {\mathbf E}}{\partial t}
\,,
\label{maxwell_norm}
\end{eqnarray}
which are completely identical to the corresponding  normalized Maxwell equations \cite{manfredi_maxw}, except for the sign in front of sources.
The following dimensionless parameters have appeared naturally in the equations:
\be
 \beta = \frac{\overline{u}}{c}, ~~~~~~ \alpha^{-1} = \frac{4\pi G \overline{\rho} L}{a}=4\pi G \overline{\rho}T^2 = \omega_J^2 T^2,
\ee
where we have defined the Jean's frequency $\omega_J=\sqrt{4\pi G \overline{\rho}}$. The parameter $\beta$ represents the reference velocity normalized to the speed of light and controls the magnitude of relativistic effects.

It is well known \cite{LBLL} that the EM Maxwell equations possess two nonrelativistic, Galilei-covariant limits, corresponding either to $| {\mathbf E}| \gg |{\mathbf B}|$ (electric limit) or $| {\mathbf E}| \ll |{\mathbf B}|$ (magnetic limit). The electric limit is recovered when $\beta \ll 1$ and $\alpha = O(1)$, and the magnetic limit when $\beta \ll 1$ and $\alpha \ll 1$, but $\alpha/\beta = O(1)$. The magnetic limit is irrelevant to GEM, because it implies the existence of at least two species of particles with opposite charge, whereas gravity is always attractive. Therefore we shall focus on the electric limit.

\paragraph{Electric limit.---}
We shall follow the method detailed in Ref. \cite{manfredi_maxw}, which consists in expanding the GEM fields in a power series of the small parameter $\beta$ (i.e., $\mathbf{E}=\mathbf{E}_0 + \beta \mathbf{E}_1 + \dots$, etc.).
To lowest (zeroth) order in $\beta$ one simply obtains the Newtonian Poisson's equation for the gravitoelectric potential $V$, i.e. $\Delta V = \rho/\alpha$.
The first correction depending on the gravitomagnetic field appears at first order in $\beta$.
Putting together the results at zeroth and first order, the GEM equations in the electric limit can be written in terms of the fields
\begin{eqnarray}
\nabla\cdot \mathbf{E} &=& -\rho/\alpha, \label{elect_lim_fields1} \\
\nabla\cdot \mathbf{B} &=& \nabla \times \mathbf{E} =0,  \\
\nabla \times \mathbf{B} &=& -\frac{\beta}{\alpha}{\mathbf J} + \beta \frac{\partial \mathbf{E}}{\partial t},
\label{elect_lim_fields3}
\end{eqnarray}
or in terms of the potentials
\begin{eqnarray}
\Delta V &=& \rho/\alpha, \label{elect_lim_pot_a}\\
\Delta \mathbf{A} &=& (\beta/\alpha) \mathbf{J}, \label{elect_lim_pot_b}
\end{eqnarray}
with the Lorentz gauge condition: $\beta \partial_t V + \nabla \cdot {\mathbf A}=0$.
Note that, in the above equations, the gravitoelectric field and potential are quantities of order zero, whereas the gravitomagnetic field (and its vector potential) are first order quantities.

For the Lorentz force, using our units and defining a reference force $\overline{\delta F} = \overline{\rho} a$, we obtain
\be
\delta\mathbf{F}= \rho\mathbf{E}+ 4\beta \mathbf{J} \times \mathbf{B}
\label{lorentz_force_norm}
\ee

\paragraph{Lorentz transformations.---}
Let us consider two reference frames traveling at relative velocity $\mathbf{v}$. Here, we shall give the Lorentz transformation to first order in $\beta$ without proof, which can be found in \cite{manfredi_maxw}.
For the space-time, we find the standard Galilean transformations:
\begin{eqnarray}
{\mathbf x}' &=& {\mathbf x}- \mathbf{v}t, \\
t' &=& t,
\end{eqnarray}
and
\begin{eqnarray}\nabla' &=& \nabla \\
\partial_{t'} &=& \partial_{t} + \mathbf{v} \cdot \nabla
\end{eqnarray}
In the electric limit, the fields transform as
\begin{eqnarray}
{\mathbf E}' &=&  {\mathbf E} \label{galilei_e}\\
{\mathbf B}' &=&  {\mathbf B} - \beta \mathbf{v} \times {\mathbf E}, \label{galilei_b}
\end{eqnarray}
and the sources
\begin{eqnarray}
\mathbf{J}' &=& \mathbf{J}-\mathbf{v} \rho, \label{galilei_j}\\
\rho' &=& \rho. \label{galilei_rho}
\end{eqnarray}
It can be easily verified that the GEM equations in the electric limit, Eqs. (\ref{elect_lim_fields1})-(\ref{elect_lim_fields3}), are left invariant by the above transformations of space-time, fields, and sources.

\paragraph{Summary.---}
In this Section, we have derived a set of reduced nonrelativistic GEM equations, valid to first order in $\beta$. They can be expressed either in terms of the potentials, as in Eqs. (\ref{elect_lim_pot_a})-(\ref{elect_lim_pot_b}), or in terms of the fields, as in Eqs. (\ref{elect_lim_fields1})-(\ref{elect_lim_fields3}). These reduced equations go beyond the Poisson equation (\ref{poisson}) of Newtonian gravity, since they also include gravitomagnetic effects. Nevertheless, they are still Gallilei covariant, as can readily be checked by applying the above Lorentz transformations to Eqs. (\ref{elect_lim_fields1})-(\ref{elect_lim_fields3}).
Such set of equations can be conveniently coupled to the (also Galilei covariant) Schr\"{o}dinger equation to construct a suitable generalization of the SNEs.

\section{Extended Schr\"{o}dinger-Newton equations}\label{sec:sne}
The nonrelativistic Hamiltonian for a {\correct scalar} particle evolving in a GEM field reads as follows
\be
H = \frac{(\mathbf{p}-2m\mathbf{A}/c)^2}{2m} + mV,
\ee
which is compatible with the Lorentz force given in Eq. (\ref{lorentz_force}).
The corresponding Schr\"{o}dinger equation is
\be
i\hbar \frac{\partial \Psi}{\partial t} = \frac{1}{2m} \left(-i\hbar \nabla-\frac{2m}{c}\mathbf{A}\right)^2 \Psi +m V \Psi.
\label{ext_sne}
\ee
If we normalize Eq. (\ref{ext_sne}) following the prescription employed in Sec. \ref{sec:gem}, we obtain:
\be
ih_0 \frac{\partial \Psi}{\partial t} = \frac{1}{2} \left(-i h_0 \nabla-4\beta\mathbf{A}\right)^2 \Psi + V \Psi,
\label{ext_sne_norm}
\ee
where $h_0 = \hbar T/(mL^2)$.
However, the standard SNEs are usually normalized using the analog of atomic units for the gravitational interaction, i.e., space is measured in units of the gravitational ``Bohr radius" $a_G =\hbar^2/(Gm^3)$ and time in units of $t_G=\hbar^3/(m^5 G^2)$.
Let us apply these units to the normalization that we employed in the preceding section for the GEM equations, i.e., let us take $L=a_G$ and $T=t_G$. In addition, we take for the mass density $\overline{\rho}=m/L^3$.
Our purpose now is to see what happens to the three dimensionless constants that we have encountered so far: $\alpha$, $\beta$, and $h_0$. A simple calculations shows that $h_0=1$.
The constant $\alpha$ becomes
\be
\alpha^{-1} = 4\pi G \overline{\rho}T^2 = 4\pi G mT^2L^{-3} = 4\pi.
\ee
Finally, for $\beta$ we obtain
\be
\beta = \frac{L}{Tc} = \frac{m^2G}{\hbar c} = \left(\frac{m}{m_P}\right)^2,
\ee
where $m_P=\sqrt{\hbar c/G}$ is the Planck mass.
As could be expected, the Planck mass appears naturally as soon as we go beyond the Newtonian approximation for the gravitational interaction.

Finally, the normalized SNEs can be written as:
\begin{eqnarray}
i\frac{\partial \Psi}{\partial t} &=& \frac{1}{2} \left(-i \nabla-4\beta\mathbf{A}\right)^2 \Psi + V \Psi, \label{sne_ext}\\
\Delta V &=&  4\pi\rho, \label{poisson_ext}\\
\Delta \mathbf{A} &=&  4\pi\beta \mathbf{J}  \label{ampere_ext}.
\end{eqnarray}
Quite naturally, for $\beta=0$ we recover the standard SNEs.
In the above equations, the (dimensionless) mass density is given by $\rho=|\Psi|^2$. The current, in our normalized units, is defined as:
\be
 \mathbf{J} = -\frac{1}{2}(\Psi \nabla\Psi^\ast - \Psi^\ast \nabla\Psi)-4\beta |\Psi|^2 \mathbf{A}.
\label{current}
\ee
However, the last term in Eq. (\ref{current}) would yield a higher-order correction to the vector potential, and can therefore be neglected.

In summary, we have derived a system of equations that generalises the standard SNEs to include corrections due to the gravitomagnetic field. Equations (\ref{sne_ext})-(\ref{ampere_ext}) constitute a minimal model where nonrelativistic (Galilean) quantum mechanics is coupled self-consistently to Galilean (but post-Newtonian) gravity. It is also a minimal model where the three universal constant $G$, $\hbar$, and $c$ appear naturally {\footnote{Even though the equations contain the constant $c$, they do not incorporate propagation at the speed of light. Indeed, since the reduced GEM equations (\ref{poisson_ext})-(\ref{ampere_ext}) are elliptic (rather than parabolic, like the wave equation) transmission of information occurs at infinite speed.}.

\paragraph{Galilean covariance.---}
The question of the Galilean covariance of the Eqs. (\ref{sne_ext})-(\ref{ampere_ext}) should be analysed more thoroughly.
As we have seen in Sec. \ref{sec:gem}, the GEM equations in the electric limit are Galilei covariant with respect to the transformation of the fields given by Eq. (\ref{galilei_e})-(\ref{galilei_b}).
The Schr\"{o}dinger equation (\ref{ext_sne}) is also Galilei covariant, but not for the same transformations. Indeed, it was shown by Brown and Holland \cite{Brown} that the Schr\"{o}dinger equation with scalar and vector potentials is Galilei covariant only when the fields and sources transform according to the Lorentz transformations in the {\em magnetic} limit:
\begin{eqnarray}
{\mathbf B}' &=&  {\mathbf B} \\
{\mathbf E}' &=&  {\mathbf E} + \beta \mathbf{v} \times {\mathbf B}. \label{galilei_m}
\end{eqnarray}
and
\begin{eqnarray}
\mathbf{J}' &=& \mathbf{J}, \label{lorentz_jrho_nondim2}\\
\rho' &=& \rho - \beta^2 \mathbf{v} \cdot \mathbf{J},
\end{eqnarray}
which differ from the transformations in the electric limit, Eqs. (\ref{galilei_e})-(\ref{galilei_rho}).
Thus the Schr\"{o}dinger equation and the fields equations are both Galilei covariant, but not under the same transformations of the fields and sources. This is a somewhat unfortunate situation, but is not different from the analogous case for the coupled Schr\"{o}dinger-Maxwell system in ordinary EM \cite{Brown}.

\paragraph{Extension to spin-1/2 particles.---}
{\correct The above results were derived for a scalar quantum particle. For a spin-1/2 particle, the spin should couple to the gravitomagnetic field in the same way as it couples to the ordinary magnetic field (albeit with a different coupling constant), as was shown by Adler et al. \cite{adler} in a context other than the SNEs. In that case, the SNEs should be replaced by a spinorial ``Pauli-Newton equation", which contains an additional Zeeman term (proportional to ${\mathbf \sigma}\cdot {\mathbf B}$) in the Hamiltonian.
Along the same lines, further relativistic effects (at second order in $1/c$) could be added using the procedure detailed in \cite{anant} for the Dirac-Maxwell equations.
}

\section{Generalization to many particles}\label{sec:manybody}
The Schr\"{o}dinger equation can be conveniently generalized to a  mixture of $N$ states by using the Wigner representation of quantum mechanics (see, for instance, \cite{Hillery}). This is based on a phase-space function defined as [we use the same notation and normalized units as in Eq. (\ref{sne_ext})]:
\be f({\mathbf r},{\mathbf p},t) = \frac{1}{2\pi}\sum_{k=1}^{N} \int_{-\infty}^{+\infty} \Psi_k^{\displaystyle*}
\left({\mathbf r} + \frac{{\mathbf \lambda}}{2},t\right) \Psi_k\left({\mathbf r} -
\frac{\mathbf{\lambda}}{2},t\right) e^{i {\mathbf p} \cdot {\mathbf \lambda}}~d\lambda
\label{wigfunc}
\ee
Such Wigner function possesses most of the properties of a true probability distribution in the phase space $({\mathbf r},{\mathbf p})$, except that it can take negative values.

The evolution equation for the Wigner function in the presence of a scalar and a vector potential is rather complicated \cite{arnold} and not particularly illuminating. However, since we know that gravitomagnetic effects are small
compared to gravitoelectric ones, it seems reasonable to neglect quantum corrections on the former and retain them only for the latter. If we further define the velocity as ${\mathbf v} = {\mathbf p}-4\beta{\mathbf A}$ (again using the normalized units of Sec. \ref{sec:sne}), the Wigner evolution equation becomes
\begin{eqnarray}
&& \frac{\partial{f}}{\partial{t}} + \mathbf{v}\cdot \nabla_{\mathbf r} f
+ 4\beta (\mathbf{v}\times \mathbf{B})\cdot \nabla_{\mathbf v} f ~+ \nonumber \frac{i}{2\pi} \times\\
&&
\int\int{d\lambda}{d{\mathbf v}'}e^{i({\mathbf v}-{\mathbf v}')\cdot \lambda}
\left[V\left({\mathbf r}+\frac{\lambda}{2},t\right)-
V\left({\mathbf r}-\frac{\lambda}{2},t\right)\right]f({\mathbf r},{\mathbf v}',t) = 0, \label{wignereq}
\end{eqnarray}
and must be coupled to the equations for the GEM fields:
\begin{eqnarray}
\Delta V &=&  4\pi \int f d {\mathbf v}, \label{poisson_wig}\\
\Delta \mathbf{A} &=&  4\pi\beta \int f {\mathbf v} d {\mathbf v} \label{ampere_wig}.
\end{eqnarray}

The Wigner function (\ref{wigfunc}) evolves in a six-dimensional phase space, which is a daunting challenge for any numerical simulation of Eq. (\ref{wignereq}). A simplified model may be obtained by assuming that the Wigner function depends only on {\em three} phase-space variables, namely one spatial co-ordinate $x$ and two velocity co-ordinates $(v_x, v_y)$. This situation corresponds to matter ``sheets" that are infinite in the $(y,z)$ plane and can flow along the $y$ direction. Such flow generates a self-consistent gravitomagnetic field directed along $z$ and a corresponding vector potential along $y$.
In this simplified geometry, Eqs. (\ref{wignereq})-(\ref{ampere_wig}) become
\begin{eqnarray}
&& \frac{\partial{f}}{\partial{t}} + v_x \frac{\partial{f}}{\partial{x}}
+ 4\beta \left(v_y B_z \frac{\partial{f}}{\partial{v_x}}- v_x B_z \frac{\partial{f}}{\partial{v_y}} \right)
+ \nonumber \frac{i}{2\pi} \times\\
&&
\int\int{d\lambda}{dv_x'}e^{i(v_x-v_x') \lambda}
\left[V\left(x+\frac{\lambda}{2},t\right)-
V\left(x-\frac{\lambda}{2},t\right)\right]f(x,v_x',v_y,t) = 0, \label{wigner_reduced}
\end{eqnarray}
\begin{eqnarray}
\partial_x^2 V &=& 4\pi \rho \equiv 4\pi \int\int f(x,v_x,v_y,t)d v_x dv_y,\\
\partial_x^2 A_y &=& 4\pi\beta J_y \equiv 4\pi\beta\int \int f(x,v_x,v_y,t) v_y  d v_x dv_y , \label{ay}
\end{eqnarray}
where $B_z = \partial_x A_y$. The above system of equations may be amenable to numerical simulations using known methods \cite{suh,jasiak}.

A further simplification is achieved by assuming that the Wigner function takes the following form:
\be
f(x,v_x,v_y,t) = g(x,v_x,t)\times \delta\left(v_y - u_y(x,t) \right),
\label{ansatz}
\ee
where $\delta$ denotes the Dirac delta. The above Ansatz is equivalent to assuming a fluid-like behavior for the flow in the $y$ direction, with all particles possessing the same velocity $u_y(x)$ at a certain point $x$. Multiplying Eq. (\ref{wigner_reduced}) by $v_y$ and integrating over $v_y$ yields the evolution equation for $J_y=\rho u_y$:
\be
\frac{\partial{J_y}}{\partial{t}} + \frac{\partial}{\partial{x}}(u J_x) + 4\beta J_x B_z =0,
\label{uy}
\ee
where $J_x = \int g v_x d v_x$.
The equation for $g$ is obtained simply by integrating Eq. (\ref{wigner_reduced}) over $v_y$:
\begin{eqnarray}
&& \frac{\partial{g}}{\partial{t}} + v_x \frac{\partial{g}}{\partial{x}}
+ 4\beta u_y B_z \frac{\partial{g}}{\partial{v_x}}
+ \nonumber \frac{i}{2\pi} \times\\
&&
\int\int{d\lambda}{dv_x'}e^{i(v_x-v_x') \lambda}
\left[V\left(x+\frac{\lambda}{2},t\right)-
V\left(x-\frac{\lambda}{2},t\right)\right]g(x,v_x',t) = 0, \label{wigner_reduced2}
\end{eqnarray}
which must be coupled to Eq. (\ref{uy}) and the equations for the potentials:
\be
\partial_x^2 V = 4\pi \int g d v_x,~~~~~~~~
\partial_x^2 A_y = 4\pi\beta J_y.
\ee
Thus, we have reduced the original six-dimensional problem to a much simpler two-dimensional problem in the phase space $(x,v_x)$, which can definitely be tackled with present computational power.

Some more physical insight can be gained by further analysing Eq. (\ref{wigner_reduced2}).
The Lorentz-force term in Eq. (\ref{wigner_reduced2}) can be written, using Eq. (\ref{ay}):
\be
4\beta u_y B_z = {1\over \rho}\frac{\partial{A_y}}{\partial{x}}
\frac{\partial^2{A_y}}{\partial x^2} = {1\over \rho} \frac{\partial}{\partial x} \left(\frac{B_z^2}{2} \right),
\ee
where one can recognize the gravitomagnetic energy density $B_z^2/2$.

It is instructive to write explicitly the first two velocity moments of Eq. (\ref{wigner_reduced2}). The zeroth-order moment is simply the continuity equation:
\be
\frac{\partial{\rho}}{\partial{t}} + \frac{\partial (u_x\rho)}{\partial x}=0,
\label{continuity}
\ee
where $u_x=J_x/\rho$. The first-order moment (equation of motion for the mean velocity $u_x$) is obtained by multiplying Eq. (\ref{wigner_reduced2})  by $v_x$ and integrating in velocity space. After some algebra, one obtains the standard hydrodynamic equation:
\be
\frac{\partial{u_x}}{\partial{t}} + u_x \frac{\partial u_x}{\partial x} = - {1\over \rho} \frac{\partial P}{\partial x} + {1\over \rho} \frac{\partial P_B}{\partial x} - \frac{\partial V}{\partial x},
\label{fluid}
\ee
where $P= \int(v_x - u_x)^2 g dv_x$ is the kinetic pressure,
$P_B=B_z^2/2$ is the ``gravitomagnetic pressure", and the last term is the gravitoelectric field.
We stress that Eq. (\ref{fluid}) originates from the fully quantum Wigner evolution Eq. (\ref{wigner_reduced2}), although it does not seem to contain any excplicitly quantum terms. These are hidden in the kinetic pressure term \cite{manfredi_hydro}, which is a second-order velocity moment that depends on the full Wigner function $g$.

From a physical point of view, Eq. (\ref{fluid}) is illuminating. It shows that our system evolves under the action of three terms: (i) the kinetic pressure $P$, which describes the usual dispersion of the wave packet; (ii) the gravitoelectric potential $V$, which, as in the standard SNEs, counteracts the dispersion and can even induce a contraction of the wave packet \cite{giulini2011,vanmeter,manfredi}; and finally (iii) the gravitomagnetic pressure $P_B$, which is in fact a negative pressure that reinforces the (attractive) gravitoelectric term.

In summary, it appears from the above example that the  gravitomagnetic correction contributes to the standard Newtonian attraction (gravitoelectric term) in counteracting the wave packet dispersion, although of course the gravitomagnetic term is much smaller.

\section{Discussion}\label{sec:discussion}
Experiments aimed at detecting the role of gravity on quantum decoherence are likely to involve the study of the interference fringes of small (micrometer) solid-state objects. These objects should be light enough to display some degree of quantum coherence, but also heavy enough to induce some measurable gravitational effects. Interferometry experiments on gold clusters \cite{nimmrichter,hornberger} are possible candidates for such studies.

For the standard SNEs, it was proven several times \cite{giulini2011,vanmeter,manfredi} that gravitational effects start affecting the Schr\"{o}dinger dynamics for masses larger than the critical mass $m_c \approx \hbar^{2/3}R^{2/3} G^{-1}$, where $R$ is the size of the object in question.
For metal clusters, the number density is fixed by their cristalline structure (for gold, $n_{\rm gold} \approx 5\times 10^{28} \rm m^{-3}$) and this determines their mass $m$ for a given size $R$. Combining this with the above relationship between $m$ and $R$, it turns out that one can expect gravitational effects (if any) to show up for metal clusters
with a size of a few microns and a mass of about $5 \times 10^9$ atomic masses \cite{manfredi}. This is not within reach of present quantum interference experiments, but it is not too far either.

Here, we have shown that semi-relativistic post-Newtonian corrections to this limit, originating from the gravitomagnetic field, are proportional to the ratio of the mass of the object to the Planck mass. In atomic mass units, this is $m_P=1.31 \times 10^{19}$ a.m.u., still ten order of magnitudes higher than the above critical mass computed for Newtonian gravity. Even for very small post-Newtonian effects (say, 1\%), the required value of the mass is far larger than what can reasonably be expected for current (and near future) interferometry experiments.

But the scope of the present work was more general. We constructed a  model where nonrelativistic quantum mechanics is coupled to semi-relativistic post-Newtonian gravity. We did so by expanding the GEM equations to the lowest order in $1/c$, but still retaining the effect of the gravitomagnetic field.
These reduced GEM equations preserve Galilei covariance.  The Schr\"{o}dinger equation with gravitoelectric and gravitomagnetic potentials is also Galilei covariant, but unfortunately not for the same transformations of the fields and potentials.

In spite of this drawback, such extended SNEs represent a minimal model where the three fundamental constants of nature ($G$, $\hbar$, and $c$) occur in a natural way. In the search for gravitational effects in mesoscopic quantum systems, this model can constitute a useful guide for future experiments and theoretical investigations.

\end{document}